\documentclass[prd,superscriptaddress,amsfonts,amssymb,amsmath,showpacs,onecolumn]{revtex4-2}
\usepackage{bm}
\usepackage{amsfonts}
\usepackage{latexsym}
\usepackage[utf8]{inputenc}
\usepackage{graphicx}
\usepackage{amsmath}
\usepackage{palatino}
\usepackage{mathpazo}
\usepackage{textcomp}
\usepackage{float}
\usepackage{booktabs}
\usepackage{dcolumn}
\usepackage{ragged2e}
\usepackage{hyperref}
\hypersetup{colorlinks,citecolor=blue}
\hypersetup{colorlinks=true,linkcolor=red,filecolor=magenta,    urlcolor=blue}
\usepackage{amsmath}
\usepackage{xcolor}
\usepackage{orcidlink}
\usepackage[caption=false]{subfig}
\usepackage{commath}
\def\jnl@style{\it}
\def\aaref@jnl#1{{\jnl@style#1}}

\def\aaref@jnl#1{{\jnl@style#1}}

\def\aj{\aaref@jnl{AJ}}                   
\def\apj{\aaref@jnl{ApJ}}                 
\def\apjl{\aaref@jnl{ApJ}}                
\def\apjs{\aaref@jnl{ApJS}}               
\def\apss{\aaref@jnl{Ap\&SS}}             
\def\aap{\aaref@jnl{A\&A}}                
\def\aapr{\aaref@jnl{A\&A~Rev.}}          
\def\aaps{\aaref@jnl{A\&AS}}              
\def\mnras{\aaref@jnl{Mon.~Not.~Roy.~Astron.~Soc.}}             
\def\prd{\aaref@jnl{Phys.~Rev.~D}}        
\def\plb{\aaref@jnl{Phys.~Lett.~B}}        
\def\prc{\aaref@jnl{Phys.~Rev.~C}}  
\def\prl{\aaref@jnl{Phys.~Rev.~Lett.}}    
\def\qjras{\aaref@jnl{QJRAS}}             
\def\skytel{\aaref@jnl{S\&T}}             
\def\ssr{\aaref@jnl{Space~Sci.~Rev.}}     
\def\zap{\aaref@jnl{ZAp}}                 
\def\nat{\aaref@jnl{Nature}}              
\def\aplett{\aaref@jnl{Astrophys.~Lett.}} 
\def\apspr{\aaref@jnl{Astrophys.~Space~Phys.~Res.}} 
\def\physrep{\aaref@jnl{Phys.~Rep.}}      
\def\physscr{\aaref@jnl{Phys.~Scr}}       
\def\commat{\aaref@jnl{Comm.~Math.~Phys.}}              
\def\science{\aaref@jnl{Science}}               
\def\cqg{\aaref@jnl{Classical Quant.~Grav.}}            
\def\jpcs{\aaref@jnl{JPCS}}                                     
\def\ijmpd{\aaref@jnl{Int.~J.~Mod.~Phys.~D}}                    
\def\grg{\aaref@jnl{Gen.~Relat.~Gravit.}}               
\def\rpp{\aaref@jnl{Rep.~Prog.~Phys.}}          
\def\npa{\aaref@jnl{Nucl.~Phys.~A}}        
\def\lrr{\aaref@jnl{Living Rev.~Rel.}}                   
\def\jcap{\aaref@jnl{J.~Cosmology Astropart.~Phys.}}    
\def\rmp{\aaref@jnl{Rev.~Mod.~Phys.}}   
\def\epjc{\aaref@jnl{Eur.~Phys.~J.~C}}

\allowdisplaybreaks[1]

\begin{document}
\color{black}       
\title{Dark-Energy Anisotropic Compact Configurations in 4D Einstein-Gauss-Bonnet Gravity: From Structure to Observational Viability}

\author{Anirudh Pradhan} 
\email[]{pradhan.anirudh@gmail.com}
\affiliation{Centre for Cosmology, Astrophysics and Space Science, GLA University, Mathura-281 406, Uttar Pradesh, India}

\author{Takol Tangphati \orcidlink{0000-0002-6818-8404}} 
\email{takoltang@gmail.com}
\affiliation{School of Science, Walailak University, Thasala, \\ Nakhon Si Thammarat, 80160, Thailand}
\affiliation{Research Center for Theoretical Simulation and Applied Research in Bioscience and Sensing, Walailak University, Thasala, Nakhon Si Thammarat 80160, Thailand}

\author{Ayan Banerjee \orcidlink{0000-0003-3422-8233}} 
\email{ayanbanerjeemath@gmail.com}
\affiliation{Astrophysics and Cosmology Research Unit, School of Mathematics, Statistics and Computer Science, University of KwaZulu--Natal, Private Bag X54001, Durban 4000, South Africa}

\author{Javlon Rayimbaev}
\email{javlon@astrin.uz}
\affiliation{Institute of Fundamental and Applied Research, National Research University TIIAME, Kori Niyoziy 39, Tashkent 100000, Uzbekistan}
\affiliation{University of Tashkent for Applied Sciences, Str. Gavhar 1, Tashkent 100149, Uzbekistan}
\affiliation{National University of Uzbekistan, Tashkent 100174, Uzbekistan}
\affiliation{Tashkent State Technical University, Tashkent 100095, Uzbekistan}
\affiliation{Urgench State University, Kh. Alimjan Str. 14, Urgench 221100, Uzbekistan}


\date{\today}

\begin{abstract}
We address the equilibrium configurations and stability properties of anisotropic compact stars whose interior is described by a modified Chaplygin gas (MCG) equation of state in the framework of the regularized four-dimensional Einstein–Gauss–Bonnet (4DEGB) theory. Applying a quasi-local prescription for the pressure anisotropy, we derive the modified Tolman–Oppenheimer–Volkoff (TOV) equations and integrate them numerically over a large parameter space in the Gauss–Bonnet coupling $\alpha$ and the degree of anisotropy $\beta$. We provide mass-radius sequences, mass-compactness, energy density, and pressure profiles, and perform a full stability analysis based on the turning-point criterion, the radial adiabatic index $\gamma_r$, and the radial and transverse sound speeds $v_r^2$ and $v_t^2$. Our results show that positive $\alpha$ and positive anisotropy $(\beta > 0)$ systematically increase the maximum mass and radius, enabling then configurations that exceed $2\,M_\odot$ while still obeying causality and the modified Buchdahl bound in 4DEGB gravity. A comparison with the latest astrophysical constraints (NICER, GW170817, GW190814, and massive-pulsar measurements) identifies regions of the $(\alpha,\beta)$ parameter space that are observationally allowable. In conclusion, anisotropic dark-energy stars in 4DEGB gravity provide viable, observationally testable ultra-compact alternatives to normal neutron stars and black holes, and also potentially open rich avenues for further multi-messenger searches for higher-curvature effects.
\end{abstract}

\maketitle

\section{Introduction}
\label{sec:intro}

The last two decades have witnessed remarkable progress in probing the nature of compact astrophysical objects through gravitational wave astronomy and precision pulsar timing. Landmark detections by LIGO/Virgo, including GW170817 from a binary neutron star merger~\cite{LIGOScientific:2018cki} and the intriguing GW190814 event involving a secondary component with mass $2.50-2.67~M_\odot$ in the so-called mass gap~\cite{LIGOScientific:2020zkf}, have opened unprecedented opportunities to constrain the equation of state of ultra-dense matter and test gravitational physics in strong-field regimes. Simultaneously, radio observations have revealed exceptionally massive pulsars such as PSR J0952-0607 ($2.35\pm0.17~M_\odot$)~\cite{Romani:2022jhd} and PSR J0740+6620 ($2.08\pm0.07~M_\odot$)~\cite{Fonseca:2021wxt}, pushing the boundaries of our understanding. While Einstein's general relativity (GR) has passed numerous tests with remarkable precision, these observations naturally prompt the question of whether modifications to GR might be needed to fully explain the diversity of compact objects we observe, particularly those in unexplored mass and density regimes.

Among the various alternatives to GR, higher curvature theories of gravity stand out as theoretically well-motivated extensions. These theories replace the linear relationship between spacetime curvature and energy-momentum found in GR with more general functions that include higher-order curvature terms. Historically, such modifications have been treated with caution due to Lovelock's theorem~\cite{Lovelock:1971yv}, which ensures that higher-order terms vanish identically in four-dimensional spacetime or less (\(D \leq 4\)). This apparent roadblock restricted higher curvature theories to dimensions \(D>4\), where they naturally emerge from string theory and other approaches to quantum gravity. The simplest nontrivial extension beyond the Einstein-Hilbert term is the Gauss-Bonnet (GB) combination, quadratic in curvature,
\begin{equation}\label{eq:gbaction}
\begin{aligned}
S_{D}^{GB} &= \alpha \int d^{D}x \, \sqrt{-g} \left[ R^{abcd} R_{abcd} - 4 R^{ab} R_{ab} + R^2 \right] \\
&\equiv \alpha \int d^{D} x \sqrt{-g} \mathcal{G},
\end{aligned}
\end{equation}
which in \(D=4\) reduces to a total derivative and contributes nothing to the dynamics.

 Recent developments have changed the landscape. Glavan and Lin~\cite{Glavan:2019inb} showed that the GB term can become dynamical in four dimensions by rescaling the coupling constant,
\begin{equation}\label{eq:alpharescale}
\alpha \rightarrow \frac{\alpha}{D-4},
\end{equation}
and then taking the limit \(D\to 4\). This prescription keeps the essential structure of the theory intact while avoiding the Lovelock restriction. The resulting theory is called novel 4DEGB gravity. It has been used to construct many solutions, including black holes with thermodynamics and geometry analysis~\cite{Ghosh:2020syx,Konoplya:2020juj,Singh:2020xju,HosseiniMansoori:2020yfj,SINGH2020100730,PhysRevD.101.104018,Yang:2020jno}, electrically charged configurations~\cite{Fernandes:2020rpa,universe8100549,Zhang:2020sjh}, and spacetimes with magnetic charge or nonlinear electrodynamics sources~\cite{Jusufi:2020qyw,ABDUJABBAROV2020100715,Jafarzade:2020ilt}. Within this framework, light deflection by black holes has been revisited~\cite{Islam:2020xmy,Jin:2020emq,Kumar:2020sag}, quasi-normal modes~\cite{Churilova:2020aca,Mishra:2020gce,Aragon:2020qdc} and black-hole shadows~\cite{Zubair:2023cep,Konoplya:2020bxa,Guo:2020zmf,Zeng:2020dco,Rayimbaev:2022znx}. Exotic compact objects like Morris–Thorne–type traversable wormholes and thin–shell constructions have also been studied~\cite{Jusufi:2020yus,Zhang:2020kxz,Chakraborty:2025cpp}.

The idea of such dimensional reduction sparked a lot of debate about the consistency of a four-dimensional formulation, with many concerns raised in the literature~\cite{Gurses:2020ofy,Ai:2020peo,Shu:2020cjw}. To overcome this limitation, two separate groups independently addressed these foundational questions~\cite{Hennigar:2020lsl,Fernandes:2020nbq}. They proceeded with the same rescaling \eqref{eq:alpharescale} initially proposed by Glavan and Lin~\cite{Glavan:2019inb} to derive consistent versions of what is now known as 4DEGB gravity. They modified the gravitational equations by adding a scalar field to the action, thus preserving the Lovelock theorem and rendering the 4DEGB a member of the Horndeski family. The first one uses conformal rescaling methods similar to the one used to obtain the $D\to 2$ dimensional limit of general relativity~\cite{Mann:1992ar}, while the second one uses the Kaluza-Klein dimensional reduction procedure~\cite{Lu:2020iav}. Despite the fact that these approaches produce equivalent theories (up to trivial field redefinitions),  there is an important distinction: the Kaluza-Klein framework introduces additional terms in the gravitational field equations that depend on the curvature of the maximally symmetric $(D-4)$-dimensional space.  When these additional contributions vanish, one arrives at the 4DEGB action contribution:
\begin{equation}\label{eq:4DEGBactionterm}
S_{4}^{GB} = \alpha \int d^4x \sqrt{-g} \left[ \phi \mathcal{G} + 4 G_{ab} \nabla^a \phi \nabla^b \phi - 4 (\nabla \phi)^2 \Box \phi + 2 (\nabla \phi)^4 \right],
\end{equation}
where the scalar field gradients are defined through $(\nabla\phi)^2\equiv g^{ab}\nabla_a\phi\nabla_b\phi$ and $(\nabla\phi)^4\equiv\big[(\nabla\phi)^2\big]^2$, with $\phi$ representing the newly introduced scalar degree of freedom. This contribution supplements the standard Einstein—Hilbert term in the complete theory, effectively modifying GR. What's remarkable is that the static, spherically symmetric black hole solutions emerging from the resulting field equations are identical to those obtained from the naive $D \to 4$ limit of $D > 4$ solutions presented in~\cite{Glavan:2019inb}, yet without ever invoking a higher-dimensional spacetime. Subsequent works have shown that 4DEGB gravity is a phenomenologically viable alternative to Einstein’s theory~\cite{Charmousis:2021npl,Clifton:2020xhc,Zanoletti:2023ori}, but the physical interpretation and observational consequences of these higher-curvature corrections are still under investigation~\cite{Fernandes:2022zrq}. Compact astrophysical objects, especially neutron stars, are the best place to test general relativity and modified gravity theories. Any viable alternative theory should be able to explain the observational properties of known compact objects and predict the gravitational wave signals from sources in the so-called mass gap, the theoretically mysterious regime between the maximum mass of neutron stars and the minimum mass of astrophysical black holes. These tight observational constraints are one of the best ways to test 4DEGB gravity against observations.

Compact objects are an ideal testbed for 4DEGB gravity and for distinguishing it from GR. After the foundational work on relativistic stars~\cite{Doneva:2020ped}, many studies have been done on neutron stars with realistic equations of state and found that positive GB coupling allows for larger maximum masses that approach the black hole mass limit, closing the mass gap~\cite{Saavedra:2024fzy, Bordbar:2024yai}. Anisotropic neutron star solutions exhibit modified mass-radius relations compared to isotropic configurations~\cite{NewtonSingh:2022rfo, Bordbar:2024yai}. Beyond neutron stars, the framework has been applied to quark stars with unified interacting equations of state~\cite{Gammon:2023uss}, strange quark stars~\cite{Banerjee:2020yhu}, electrically charged quark stars~\cite{Pretel:2021czp}, color-flavor locked strange stars~\cite{Banerjee:2020stc}, and white dwarfs showing deviations from the Chandrasekhar limit~\cite{Pretel:2025roz}. Complementary studies have explored barotropic equations of state in Einstein-Maxwell-GB gravity~\cite{Hansraj:2024qbe} and gravitational collapse dynamics ~\cite{Jaryal:2022rzd}. These investigations demonstrate that 4DEGB gravity yields rich phenomenology with testable predictions distinguishing it from GR.

However, while recent studies have extended 4DEGB compact star investigations beyond conventional neutron stars to include quark matter and white dwarfs, all existing configurations share a common feature: they rely on equations of state with positive or vanishing effective pressures throughout the stellar interior. An alternative, largely unexplored possibility is that some compact objects might be composed of exotic matter with unusual thermodynamic properties, including regions of negative effective pressure. One particularly interesting candidate is the class of models based on the Chaplygin gas and its generalizations. Originally introduced to unify dark matter and dark energy in cosmology~\cite{Kamenshchik:2001cp,Bilic:2001cg}, the Chaplygin gas exhibits a remarkable property: it interpolates between pressureless dust at high densities and a negative-pressure dark energy component at low densities. The modified Chaplygin gas (MCG), developed to improve agreement with observational data~\cite{Saadat_2012,Sen:2005sk,Chimento:2003ta}, extends this behavior by introducing additional parameters that allow for a more flexible matter-energy crossover. This equation of state has been extensively studied in cosmological contexts~\cite{Xu_2012,Thakur_2009,Pourhassan:2013sw}, where it provides a unified description of the cosmic acceleration without invoking separate dark matter and dark energy components.
More recently, the MCG has been applied to model stellar interiors in GR and other modified theories~\cite{BagheriTudeshki:2023dbm,Pretel:2024tjw,Jyothilakshmi:2024zqn,Das:2024ugy,Pretel:2024vvt,Das:2025lze,Banerjee:2025zhp}, giving rise to so-called ``dark energy stars''—compactobjects supported by exotic equations of state with negative effective pressures. These configurations represent intriguing alternatives to conventional compact stars, potentially explaining objects with unusual compactness or mass-radius characteristics. Despite this rich phenomenology, dark energy stars have not yet been investigated in the context of 4DEGB gravity.

A further layer of complexity in modeling compact stars arises from pressure anisotropy. Under the extreme conditions inside dense stellar objects—densities exceeding nuclear saturation, strong magnetic fields, phase transitions, or the presence of exotic matter—there is no fundamental reason to expect the radial and tangential pressures to be equal. The difference $\Delta = p_t-p_r$ can arise from a variety of physical mechanisms, including pion condensation, type-3A superfluidity, solid core formation, or relativistic kinetic effects~\cite{Heiselberg:1999mq,Zeldovich:1961sbr,Sawyer:1972cq,Herrera:1997plx}. Incorporating anisotropy into stellar models is essential for realistic descriptions, as it significantly affects the hydrostatic equilibrium structure, the maximum achievable mass, and dynamical stability. Among the various prescriptions for modeling anisotropy, the quasi-local approach introduced by Horvat and collaborators~\cite{Horvat:2010xf} has proven particularly useful. This phenomenological framework relates the anisotropy magnitude to the local compactness, naturally ensuring that the anisotropy vanishes at the stellar center and increases toward the surface in a physically reasonable way. Anisotropic compact star models have been studied extensively in GR~\cite{Maurya:2018kxg,Raposo:2018rjn,Doneva:2012rd,Mondal:2023wwo,Suarez-Urango:2023ikq} and, more recently, in modified theories including $R^2$ gravity and $f(R,T)$ theory~\cite{Folomeev:2018ioy,Naz:2024yua,Errehymy:2024tqr}. Nevertheless, the combined effects of anisotropy and exotic dark energy equations of state in 4DEGB gravity remain completely unexplored.

This gap in the literature motivates the present investigation. While neutron stars with conventional nuclear matter have been studied in 4DEGB gravity~\cite{Bordbar:2024yai,Saavedra:2024fzy}, and dark energy stars with Chaplygin-type equations of state have been examined in GR and other frameworks~\cite{Jyothilakshmi:2024zso,Tudeshki:2023ias,Panotopoulos:2020kgl}, \textit{the combined scenario of anisotropic dark energy stars in 4D Einstein-GB gravity has not been addressed}. Furthermore, the intricate interplay between the GB coupling parameter $\alpha$, the pressure anisotropy parameter $\beta$, and the parameters of the modified Chaplygin gas equation of state—and how these jointly affect stellar structure, stability, and observational viability—remains an open question. Understanding this multidimensional parameter space is crucial for assessing whether such exotic configurations can match the properties of observed compact objects, particularly those in unusual mass or compactness regimes.

In this work, we address these open questions by constructing and analyzing equilibrium models of anisotropic dark energy stars in 4DEGB gravity. Our investigation makes several novel contributions. First, we present the \textit{first systematic study} of dark energy stars in the 4DEGB framework, employing the modified Chaplygin gas equation of state to describe the stellar interior. Second, we incorporate pressure anisotropy via the quasi-local prescription of Horvat~\cite{Horvat:2010xf}, thereby extending beyond the isotropic approximation and capturing more realistic internal dynamics. Third, we perform a comprehensive stability analysis via radial perturbation theory, computing fundamental mode oscillation frequencies to delineate stable and unstable equilibrium branches in the $(\alpha,\beta)$ parameter space. Finally, we systematically explore how variations in the GB coupling $\alpha$ and anisotropy parameter $\beta$ affect key observables such as maximum mass, radius, surface redshift, and compactness. We confront our theoretical predictions with current astrophysical constraints from massive pulsars (PSR J0952-0607, PSR J0740+6620, PSR J0348+0432) and gravitational wave events (GW170817, GW190814).

The remainder of this paper is organized as follows. In Sec.~\ref{sec:4DEGB}, we review the formulation of 4DEGB gravity and derive the modified field equations for a spherically symmetric, static spacetime. Section~\ref{sec:TOV} presents the generalized Tolman-Oppenheimer-Volkoff (TOV) equations incorporating both the 4DEGB corrections and pressure anisotropy. In Sec.~\ref{sec:EOS}, we describe the modified Chaplygin gas equation of state and the quasi-local anisotropy model used in our analysis. Section~\ref{sec:results} details our numerical methods and presents the main results, including mass-radius diagrams, compactness profiles, energy density and pressure distributions, and stability analysis through radial oscillation frequencies. In Sec.~\ref{sec:stability}, we compare our theoretical predictions with observational data from massive pulsars and gravitational-wave detections, and discuss the parameter regions consistent with current constraints. Finally, Sec.~\ref{sec:conclusions} summarizes our findings, discusses their implications, and outlines directions for future work.

\section{4D Einstein-GB Gravity}\label{sec:4DEGB}

We adopt the 4DEGB gravity to study anisotropic dark energy stars. This theory extends general relativity by adding higher-curvature corrections through the GB invariant coupled to a scalar field. Following Ref.~\cite{Hennigar:2020lsl}, the action is given by:
\begin{equation}\label{eq:4DEGBaction}
S_{\text{EGB}} = \frac{1}{2\kappa} \int d^4x \sqrt{-g} \left[ R - 2\Lambda + \alpha \left( \phi \mathcal{G} + 4G_{ab}\nabla^a \phi \nabla^b \phi - 4(\nabla \phi)^2 \Box \phi + 2((\nabla \phi)^2)^2 \right) \right] + S_m,
\end{equation}
where $\kappa = 8\pi G c^{-4}$ is the gravitational coupling and $\Lambda$ is the cosmological constant. The parameter $\alpha$ represents the GB coupling with dimensions of length squared. Here $\phi$ is a dimensionless scalar field and $S_m$ is the matter action. The GB invariant is defined as:
\begin{equation}\label{eq:GBinvariant}
\mathcal{G} = R_{abcd}R^{abcd} - 4R_{ab}R^{ab} + R^2.
\end{equation}

The action (\ref{eq:4DEGBaction}) exhibits shift symmetry under:
\begin{equation}\label{eq:shiftsymm}
\phi \rightarrow \phi + C,
\end{equation}
where $C$ is any constant. This symmetry eliminates certain pathologies and ensures that initial value problems are well-defined.

Varying the action with respect to $\phi$ yields \cite{Hennigar:2020lsl}:
\begin{equation}\label{eq:phivariation}
\begin{split}
\mathcal{G} - 8G_{ab}\nabla^a \nabla^b \phi - 8R_{ab}\nabla^a \phi \nabla^b \phi + 8(\Box \phi)^2 - 8\nabla_a \nabla_b \phi \nabla^a \nabla^b \phi \\
- 16\nabla_a \nabla_b \phi \nabla^b \phi \nabla^a \phi - 8(\nabla \phi)^2 \Box \phi = 0.
\end{split}
\end{equation}

Variation with respect to the metric gives the modified field equations:
\begin{equation}\label{eq:fieldequations}
\begin{split}
G_{ab} + \Lambda g_{ab} + \alpha \bigg[ \phi H_{ab} - 2R[\nabla_a \nabla_b \phi + (\nabla_a \phi)(\nabla_b \phi)] + 8R^{c}_{\  (a} \nabla_{b)} \nabla_c \phi \\
+ 8R^{c}_{\  (a} \nabla_{b)} \phi \nabla_c \phi - 2G_{ab}[(\nabla \phi)^2 + 2\Box \phi] - 4[\nabla_a \nabla_b \phi + (\nabla_a \phi)(\nabla_b \phi)]\Box \phi \\
- [g_{ab}(\nabla \phi)^2 - 4(\nabla_a \phi)(\nabla_b \phi)](\nabla \phi)^2 + 8\nabla_c \nabla_{(a} \phi (\nabla_{b)} \phi)\nabla^c \phi \\
- 4g_{ab} R^{cd}[\nabla_d \nabla_c \phi + (\nabla_d \phi)(\nabla_c \phi)] + 2g_{ab}(\Box \phi)^2 \\
- 2g_{ab}(\nabla_c \nabla_d \phi)(\nabla^c \nabla^d \phi) - 4g_{ab}(\nabla_c \nabla_d \phi)(\nabla^c \phi)(\nabla^d \phi) \\
+ 4(\nabla_a \nabla_c \phi)(\nabla_b \nabla^c \phi) + 4R_{acbd}[\nabla^c \nabla^d \phi + (\nabla^c \phi)(\nabla^d \phi)] \bigg] = \frac{8\pi G}{c^4} T_{ab},
\end{split}
\end{equation}
where $T_{ab}$ is the stress-energy tensor:
\begin{equation}\label{eq:stresstensor}
T_{ab} := -\frac{2}{\sqrt{-g}} \frac{\delta S_m}{\delta g^{ab}}.
\end{equation}

The tensor $H_{ab}$ in Eq.~(\ref{eq:fieldequations}) is constructed from Riemann curvature contractions:
\begin{equation}\label{eq:Hab}
H_{ab} = 2R_{acde}R_{b}^{\ cde} - 4R_{ac}R_{b}^{\ c} - 4R_{acbd}R^{cd} + 2RR_{ab} - \frac{1}{2}\mathcal{G}g_{ab}.
\end{equation}

This tensor vanishes identically in four dimensions due to topological properties of $\mathcal{G}$. In what follows, we set $\Lambda = 0$ for compact star modeling in asymptotically flat spacetime. This framework enables us to derive the modified TOV equations governing anisotropic dark energy stars in 4DEGB gravity.

\subsection{Equilibrium configurations via modified TOV equations}\label{sec:TOV}

To describe the equilibrium structure of anisotropic dark-energy compact stars in 4DEGB gravity, we derive modified stellar-structure equations by incorporating curvature corrections from the GB term into the classical Tolman–Oppenheimer–Volkoff framework. We consider a static, spherically symmetric stellar configuration described by the metric ansatz
\begin{equation}\label{metric}
ds^{2} = - e^{2\Phi(r)} \, dt^{2} + e^{2\Psi(r)} \, dr^{2} + r^{2}\left( d\theta^{2} + \sin^{2}\theta \, d\varphi^{2} \right),
\end{equation}
where the radial functions $\Phi(r)$ and $\Psi(r)$ encode the gravitational field structure throughout the stellar interior and exterior. The material content is modeled as an anisotropic dark-energy fluid with stress-energy tensor
\begin{equation}\label{EMT}
T_{\mu\nu}= (\rho+p_{t})u_{\mu}u_{\nu}+p_{t}g_{\mu\nu}-\sigma\chi_{\mu}\chi_{\nu},
\end{equation}
where $\rho$ denotes the energy density, $p_r$ and $p_{t}$ represent the radial and tangential pressures respectively, and the anisotropy is quantified by $\sigma \equiv p_{t} - p_r$. The four-velocity $u^\mu$ and the unit radial vector $\chi^\mu$ satisfy the normalization conditions $u_\mu u^\mu = -1$, $\chi_\mu \chi^\mu = 1$, and orthogonality $u^\mu \chi_\mu = 0$.

Substituting the metric (\ref{metric}) and the anisotropic energy-momentum tensor (\ref{EMT}) into the 4DEGB field equations (\ref{eq:fieldequations}), we derive the temporal ($tt$) and radial ($rr$) components. Following the methodology established in previous studies of compact objects in 4DEGB gravity \cite{Doneva:2020ped,Saavedra:2024fzy,Banerjee:2020yhu}, these field equations assume the compact forms
\begin{align}
\frac{2}{r}\!\left(1+\frac{2\alpha(1-e^{-2\Psi})}{r^{2}}\right)\!\frac{d\Psi}{dr} &=
e^{2\Psi}\!\left[8\pi\rho
- \frac{1-e^{-2\Psi}}{r^{2}}\!\left(1-\frac{\alpha(1-e^{-2\Psi})}{r^{2}}\right)\right], \label{eq:FE1}\\[4pt]
\frac{2}{r}\!\left(1+\frac{2\alpha(1-e^{-2\Psi})}{r^{2}}\right)\!\frac{d\Phi}{dr} &=
e^{2\Psi}\!\left[8\pi p_r
+ \frac{1-e^{-2\Psi}}{r^{2}}\!\left(1-\frac{\alpha(1-e^{-2\Psi})}{r^{2}}\right)\right]. \label{eq:FE2}
\end{align}
The requirement of energy–momentum conservation, $\nabla^\mu T_{\mu\nu} = 0$, yields the hydrostatic equilibrium condition
\begin{equation}\label{ConsEq}
p_{r}' = -(\rho+p_{r})\Phi' + \frac{2\sigma}{r},
\end{equation}
which governs the radial pressure gradient in the presence of anisotropic stresses.

Within the 4DEGB framework, the metric potential $\Psi(r)$ relates to the enclosed gravitational mass $m(r)$ through
\begin{equation}\label{PsiEq}
e^{-2\Psi}=1+\frac{r^{2}}{2\alpha}
\Bigl[1-\sqrt{1+\frac{8\alpha m(r)}{r^{3}}}\Bigr].
\end{equation}
One can verify that as the GB coupling parameter vanishes ($\alpha\to0$), this expression correctly reduces to the standard Schwarzschild interior form $e^{-2\Psi}=1-2m/r+\mathcal{O}(\alpha)$, thereby recovering general relativity in the appropriate limit. By eliminating $\Psi$ between Eqs.~\eqref{eq:FE2} and \eqref{PsiEq}, we obtain an explicit expression for the gravitational potential gradient
\begin{equation} \label{eq:Phiprime}
\Phi'=\frac{r^{3}\bigl(\sqrt{1+8\alpha m/r^{3}}
+8\pi\alpha p_{r}-1\bigr)-2\alpha m}
{r^{2}\bigl[(r^{2}+2\alpha)\sqrt{1+8\alpha m/r^{3}}
-r^{2}-8\alpha m/r\bigr]}.
\end{equation}
Upon substituting this result into Eqs.~\eqref{eq:FE1} and \eqref{ConsEq}, we arrive at the modified TOV system for anisotropic dark-energy stars in 4DEGB gravity:
\begin{align}
m' &= 4\pi r^{2}\rho,\label{TOV1}\\[1mm]
p_{r}' &= -\frac{(\rho+p_{r})\bigl[2\alpha m
+r^{3}(1-\mathcal{A}-8\pi\alpha p_{r})\bigr]}
{r^{2}\mathcal{A}\bigl(r^{2}+2\alpha-r^{2}\mathcal{A}\bigr)}
+\frac{2\sigma}{r},\label{TOV2}
\end{align}
where we have introduced the abbreviation $\mathcal{A}\equiv\sqrt{1+8\alpha m/r^{3}}$ for notational convenience. In the dual limits $\alpha\!\to\!0$ and $\sigma\!\to\!0$, these equations manifestly reduce to the familiar isotropic TOV equations of Einstein's general relativity.

To close this system of differential equations, we adopt a modified Chaplygin gas equation of state describing the dark-energy component, $p_r=p_r(\rho)$, along with a quasi-local anisotropy prescription $\sigma=\sigma(p_r,\mu)$, where the compactness function $\mu(r)\equiv 2m(r)/r$ characterizes the local gravitational field strength. We impose regularity at the stellar center by requiring $\sigma(0)=0$, and demanding that the radial and tangential sound speed satisfy causality constraints throughout the stellar interior ($0 \leq dp_{r,t}/d\rho \leq 1$). These closure relations, combined with appropriate boundary conditions at the center and surface, render the system determinate and allow numerical integration to construct equilibrium sequences of anisotropic dark-energy stars in 4DEGB gravity.

\section{EoS and Anisotropy Model}\label{sec:EOS}

\subsection{Modified Chaplygin Gas Equation of State}\label{subsec:MCG}

To model the dark-energy component within our stellar configuration, we adopt the modified Chaplygin gas (MCG) equation of state, which has been extensively employed to describe the unification of dark matter and dark energy in both cosmological and astrophysical contexts~\cite{Kamenshchik:2001cp,Bilic:2001cg}. The MCG prescription provides a phenomenological framework that interpolates between a matter-dominated regime at high densities and a dark-energy-dominated regime at low densities, making it particularly suitable for modeling compact stellar objects with exotic interiors~\cite{Pretel:2024tjw,Jyothilakshmi:2024zqn}.

The radial pressure $p_r$ as a function of energy density $\rho$ is given by:
\begin{equation}
p_r(\rho) = A^2 \rho - \frac{B^2}{\rho},
\end{equation}
where $A$ is a dimensionless positive constant, $B$ is a positive constant with dimensions of energy density, and the linear term $A^2\rho$ represents a barotropic fluid contribution. In contrast, the negative term $-B^2/\rho$ encodes the exotic dark-energy behavior characterized by negative effective pressure~\cite{Bilic:2001cg,Sen:2005sk}. An essential physical requirement for any realistic equation of state is the causality condition, which demands that the radial sound speed $v_{sr}$ must not exceed the speed of light throughout the stellar interior~\cite{Pretel:2024tjw,Pretel:2023nhf}. The squared radial sound speed is defined as:
\begin{equation}
v_{sr}^2 = \frac{dp_r}{d\rho} = A^2 + \frac{B^2}{\rho^2}.
\end{equation}
For our numerical investigations, we adopt parameter values consistent with previous studies of dark-energy stars while ensuring compliance with the causality constraint~\cite{Jyothilakshmi:2024zqn,Das:2024ugy}.

\begin{figure}
    \centering
    \includegraphics[width = 8.5 cm]{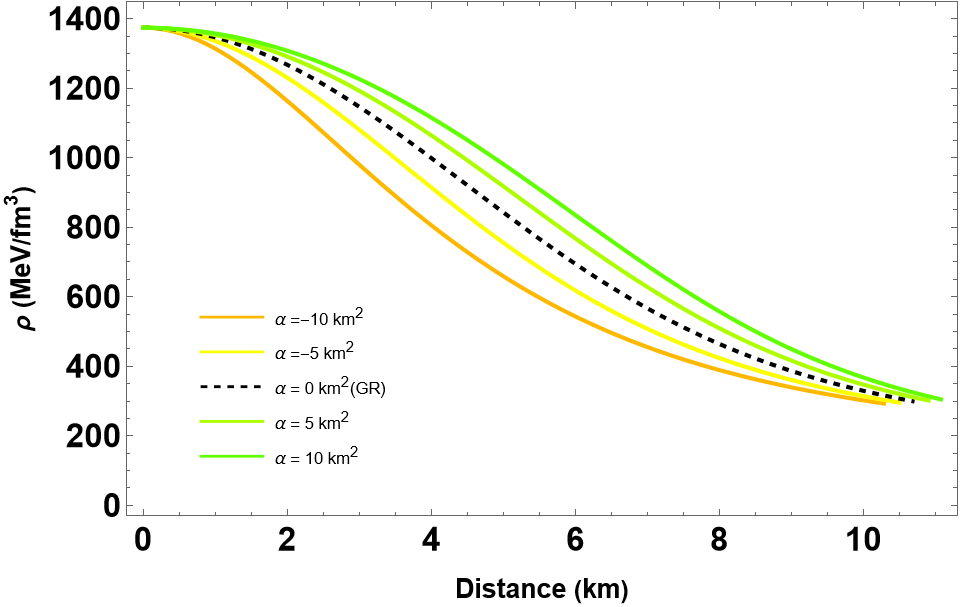}
    \includegraphics[width = 8.5 cm]{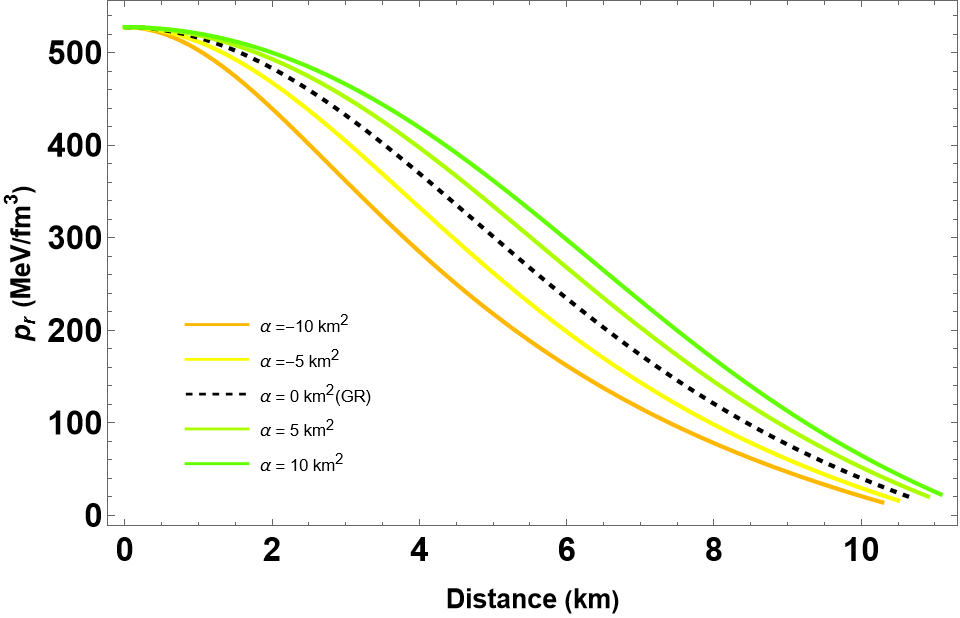} 
     \caption{These profiles are the energy density $\rho(r)$ (left panel) and radial pressure $p_r(r)$ (right panel) for anisotropic dark-energy stars in 4D Einstein–Gauss–Bonnet gravity with different values of the Gauss–Bonnet coupling parameter $\alpha$ in the range $[-10, +10]~\mathrm{km^2}$. The numerical integration is performed using the parameter set summarized in Table~\ref{table1}, with the equation of state constants fixed at $A = \sqrt{0.4}$, $B = 0.23 \times 10^{-3}~\mathrm{km^{-2}}$, and $\beta = 0.5$. The solutions remain regular throughout the stellar interior, exhibit finite central values and monotonic radial behavior, and satisfy all physically acceptable energy conditions.
}
 \label{fig1}
\end{figure}

\begin{figure}
    \centering
    \includegraphics[width = 8.5 cm]{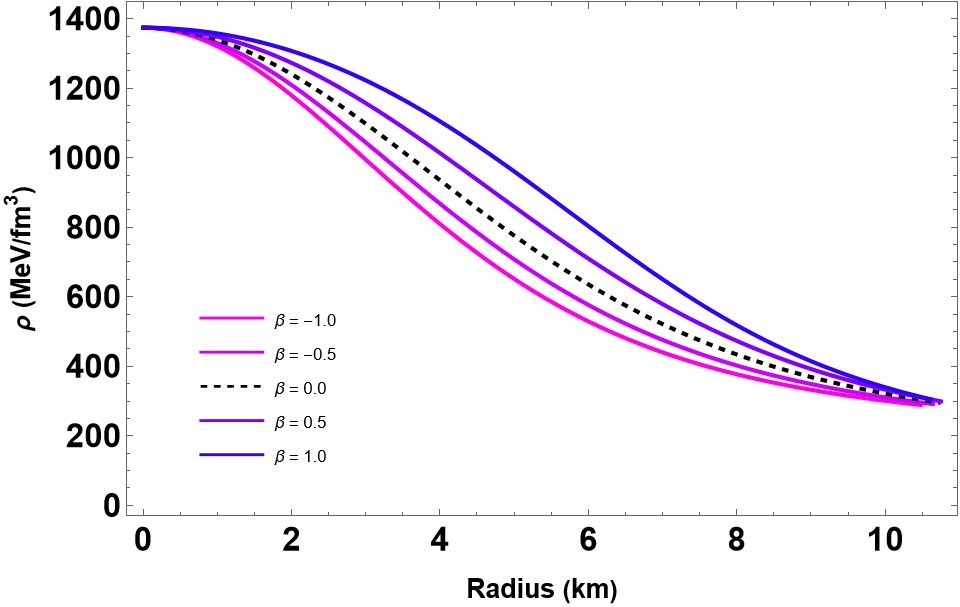}
    \includegraphics[width = 8.5 cm]{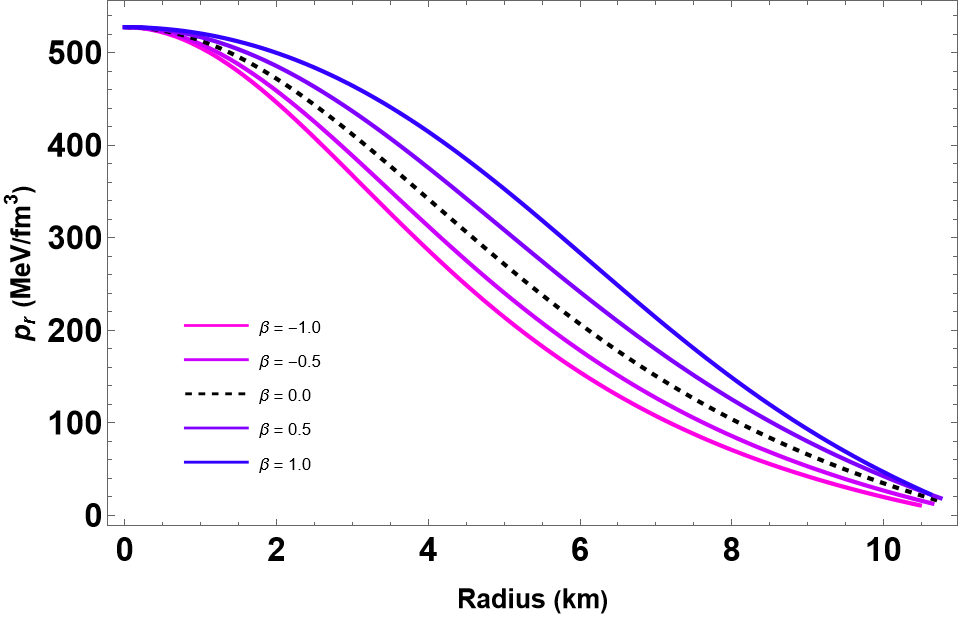} 
    \caption{These profiles are the energy density $\rho(r)$ (left panel) and radial pressure $p_r(r)$ (right panel) for anisotropic dark-energy stars in 4D Einstein–Gauss–Bonnet gravity with different values of the anisotropy parameter $\beta$ in the range $[-1.0, +1.0]$. The numerical solutions are obtained using the same parameter set summarized in Table~\ref{table2}, where the equation of state constants are fixed at $A = \sqrt{0.4}$, $B = 0.23 \times 10^{-3}~\mathrm{km^{-2}}$, and $\alpha = 5~\mathrm{km^2}$. The case $\beta = 0$ corresponds to the isotropic configuration, while negative and positive values of $\beta$ represent increasingly anisotropic distributions with dominant radial and tangential pressures, respectively. The solutions remain regular throughout the stellar interior, exhibit finite central values and monotonic radial behavior, and satisfy all physically acceptable energy conditions.
}
 \label{fig2}
\end{figure}

\subsection{Incorporating Anisotropic Pressure with the Quasi-Local Model}\label{subsec:QL_anisotropy}

Pressure anisotropy---where the tangential pressure $p_t$ differs from the radial pressure $p_r$---can arise in compact stars through various physical mechanisms, including strong magnetic fields, phase transitions, relativistic kinetic effects, and the presence of exotic matter~\cite{Herrera:1997plx,Raposo:2018rjn}. In the context of dark-energy stars, anisotropic stresses may be induced by the interplay between the modified Chaplygin gas component and the strong gravitational field corrections introduced by 4D Einstein--GB gravity.

To incorporate anisotropic effects into our stellar model, we employ the quasi-local (QL) anisotropy prescription proposed by Horvat and collaborators~\cite{Horvat:2010xf}. This model provides a phenomenological closure relation that links the anisotropy magnitude $\sigma \equiv p_t - p_r$ to the local gravitational field strength. The QL ansatz is expressed as
\begin{equation}\label{eq:aniso}
\sigma(r) = \beta p_r(r) \mu(r),
\end{equation}
where $\beta$ is a dimensionless free parameter quantifying the anisotropy strength, and $\mu(r) = 2m(r)/r$ represents the local compactness function. The parameter $\beta$ can assume both positive and negative values: positive (negative) $\beta$ corresponds to tangentially (radially) dominated pressure anisotropy, i.e., $p_t > p_r$ ($p_t < p_r$). Based on previous investigations on anisotropic compact stars in general relativity and modified gravity theories, typical values of $\beta$ lie within the range $|\beta| \lesssim 2$~\cite{Folomeev:2018ioy,Doneva:2012rd,Bordbar:2024yai}.

The QL model possesses several desirable mathematical and physical properties. First, the anisotropy vanishes identically at the stellar center, $\sigma(0) = 0$, since $\mu(0) = 0$, thereby ensuring regularity of all physical quantities at $r = 0$. Second, the anisotropy naturally approaches zero at the stellar surface $r = R$, where both $p_r(R) = 0$ and $p_t(R) = 0$ are satisfied, satisfying the junction conditions for matching the interior solution to the exterior vacuum. Third, unlike alternative anisotropy models such as the Bowers--Liang prescription~\cite{Bowers:1974tgi}, the QL ansatz ensures that anisotropic contributions vanish in the Newtonian limit, which is a physically appealing feature. These properties make the QL model particularly well-suited for constructing self-consistent equilibrium configurations of anisotropic dark-energy stars~\cite{Horvat:2010xf}, and we extend its application to the 4DEGB gravity framework in this work.

Substituting Eq.~\eqref{eq:aniso} into the modified TOV equation~\eqref{TOV2}, the hydrostatic equilibrium equation becomes
\begin{equation}
p_r' = -\frac{(\rho + p_r)\left[2\alpha m + r^3(1-\mathcal{A}-8\pi\alpha p_r)\right]}{r^2 \mathcal{A}\left(r^2 + 2\alpha - r^2 \mathcal{A}\right)} + \frac{2\beta p_r \mu}{r}.
\end{equation}
The anisotropy term $2\beta p_r \mu/r$ explicitly modifies the pressure gradient, thereby altering the mass-radius relation, compactness, and stability properties of the resulting stellar configurations. By varying $\beta$ while keeping the GB coupling $\alpha$ and the MCG parameters $(A,B)$ fixed, we can systematically investigate the impact of anisotropic stresses on the macroscopic properties of dark-energy stars in 4DEGB gravity.

\begin{figure}
    \centering
    \includegraphics[width = 8.5 cm]{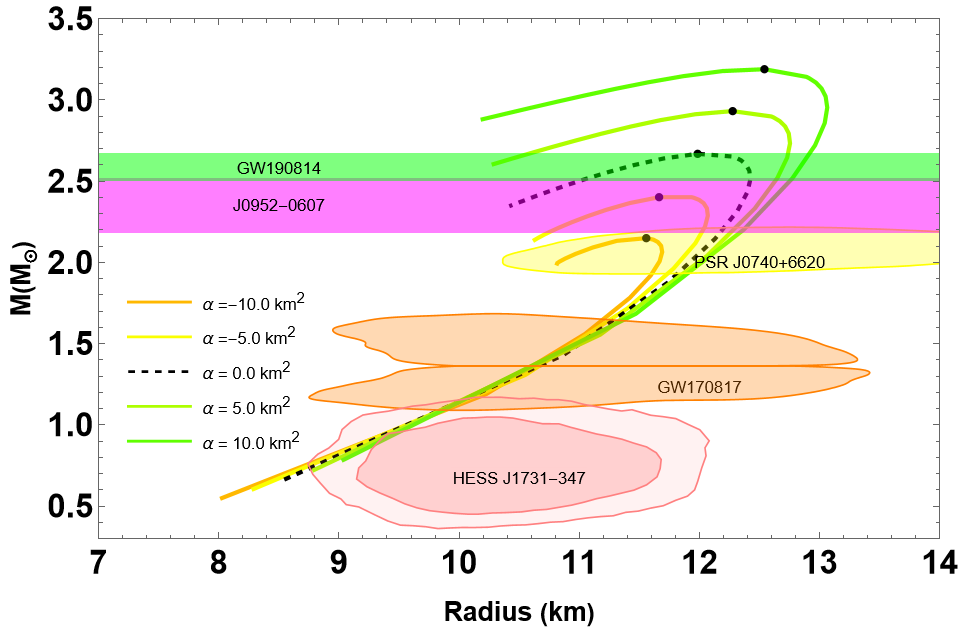}
    \includegraphics[width = 8.5 cm]{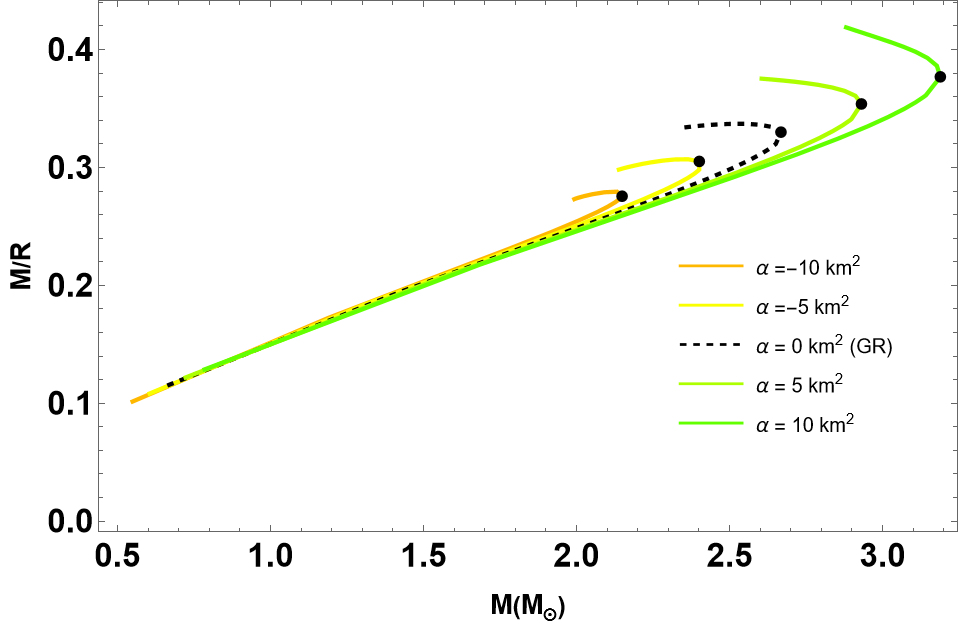} 
\caption{Mass-radius (left panel) and mass versus compactness (right panel) relations for compact stars in 4DEGB gravity with varying coupling parameter $\alpha$ (ranging from $-10$ km$^2$ to $+10$ km$^2$). The equation of state and model parameters are identical to those adopted in Figure~\ref{fig1}. The left panel includes observational constraints from massive pulsars PSR J0952$-$0607 with $M = 2.35 \pm 0.17\,M_{\odot}$ \cite{Romani:2022jhd}, PSR J0740+6620 with $M = 2.08^{+0.07}_{-0.07}\,M_{\odot}$ \cite{Fonseca:2021wxt}, PSR J0348+0432 with $M = 2.01 \pm 0.04\,M_{\odot}$ \cite{Antoniadis:2013pzd}, and the low-mass compact object in HESS J1731$-$347 \cite{Doroshenko:2022nwp}. Gravitational wave constraints from GW170817 \cite{LIGOScientific:2018cki} and the secondary component of GW190814 with mass $2.50$--$2.67\,M_{\odot}$ \cite{LIGOScientific:2020zkf} are also shown. Different values of $\alpha$ demonstrate how the GB coupling modifies the stellar structure, with the General Relativity (GR) limit corresponding to $\alpha = 0$ km$^2$. 
}
 \label{fig3}
\end{figure}

\begin{figure}
    \centering
    \includegraphics[width = 8.5 cm]{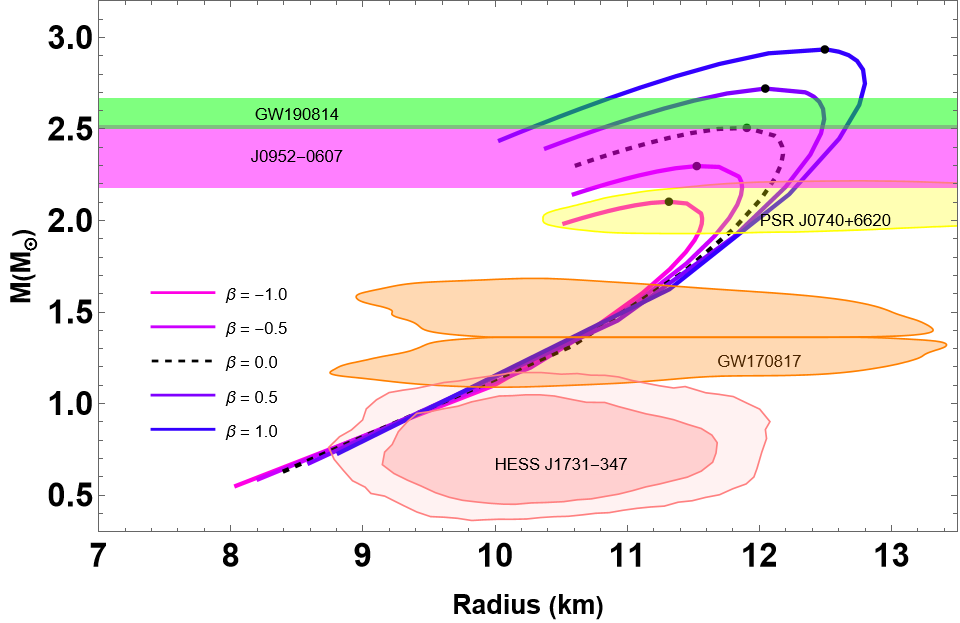}
    \includegraphics[width = 8.5 cm]{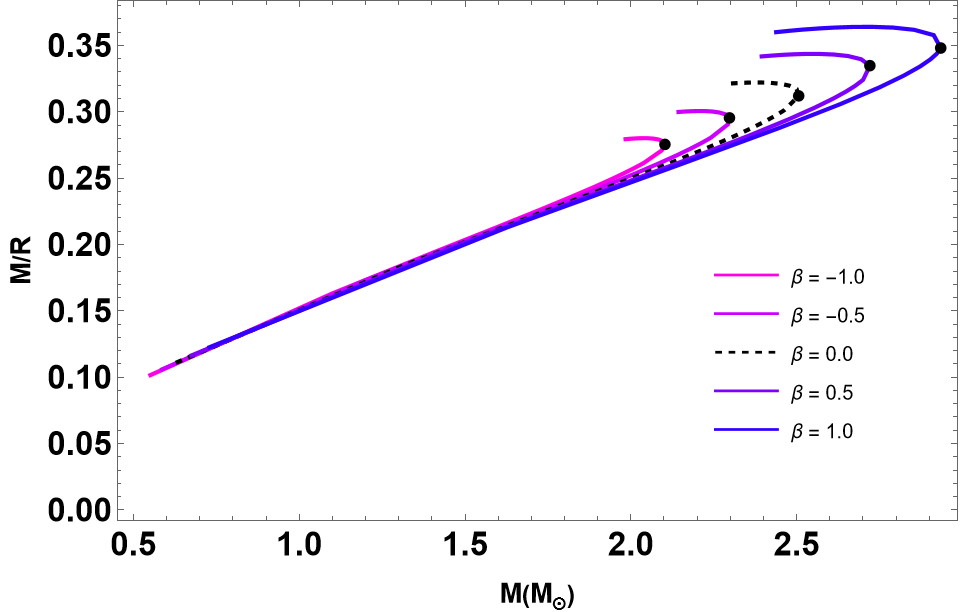} 
    \caption{Mass-radius (left panel) and mass versus compactness (right panel) relations for isotropic and anisotropic compact stars in 4DEGB gravity with varying anisotropy parameter $\beta$ (ranging from $-1.0$ to $+1.0$). The equation of state and model parameters are identical to those adopted in Figure~\ref{fig2}, while the observational constraints are from Fig.~\ref{fig3}.  The case $\beta = 0$ corresponds to the isotropic configuration, while negative and positive values of $\beta$ represent different degrees of pressure anisotropy.}
 \label{fig4}
\end{figure}


\begin{table}[htbp]
\centering
\caption{Properties of the maximum-mass configurations for compact stars in 4DEGB gravity for selected values of the GB coupling constant $\alpha$ (in km$^{2}$). Listed are the maximum gravitational mass $M_{\mathrm{max}}$ (in $M_{\odot}$), the corresponding radius $R_{\mathrm{max}}$ (in km), the central energy density $\rho_{c}$ (in MeV/fm$^{3}$), and the compactness $M_{\mathrm{max}}/R_{\mathrm{max}}$. All configurations are obtained using the same equation-of-state parameters: $A=\sqrt{0.4}$, $B=0.23\times10^{-3}~\mathrm{km^{-2}}$, and $\beta=0.5$.}
\vspace{0.3em}
\begin{tabular}{cccccc}
\hline\hline
$\alpha$ [km$^{2}$] & $M_{\mathrm{max}}$ [$M_{\odot}$] & $R_{\mathrm{max}}$ [km] & $\rho_{c}$ [MeV/fm$^{3}$] & $M_{\mathrm{max}}/R_{\mathrm{max}}$ \\
\hline
$-10$ & $2.15$ & $11.56$ & $701$ & $0.276$ \\
$-5$  & $2.40$ & $11.66$ & $905$ & $0.305$ \\
$0$   & $2.67$ & $11.98$ & $905$ & $0.330$ \\
$5$   & $2.93$ & $12.28$ & $905$ & $0.354$ \\
$10$  & $3.19$ & $12.54$ & $905$ & $0.377$ \\
\hline\hline
\end{tabular}
\label{table1}
\end{table}

\begin{table}[h]
\centering
\caption{Properties of the maximum-mass configurations for isotropic and anisotropic compact stars in 4DEGB gravity for selected values of the anisotropy parameter $\beta$. Listed are the maximum gravitational mass $M_{\text{max}}$ (in $M_{\odot}$), the radius $R_{\text{max}}$ (in km) at maximum mass, the central energy density $\rho_c$ (in MeV/fm$^3$), and the compactness $M_{\text{max}}/R_{\text{max}}$. All configurations adopt the same equation of state parameters: $A = \sqrt{0.4}$, $B = 0.23 \times 10^{-3}$ km$^{-2}$, and $\alpha = 1.0$ km$^2$. The case $\beta = 0$ corresponds to the isotropic configuration.}
\label{table2}
\begin{tabular}{ccccc}
\hline\hline
$\beta$ & $M_{\text{max}}$ [$M_{\odot}$] & $R_{\text{max}}$ [km] & $\rho_c$ [MeV/fm$^3$] & $M_{\text{max}}/R_{\text{max}}$ \\
\hline
$-1.0$ & 2.10 & 11.32 & 1110 & 0.276 \\
$-0.5$ & 2.30 & 11.53 & 1110 & 0.295 \\
$0.0$ & 2.51 & 11.91 & 905 & 0.312 \\
$0.5$ & 2.72 & 12.05 & 905 & 0.335 \\
$1.0$ & 2.93 & 12.50 & 701 & 0.348 \\
\hline\hline
\end{tabular}
\end{table}

\section{Numerical Implementation and Computational Setup}\label{sec:results}

We construct equilibrium configurations by numerically solving the modified TOV equations~\eqref{TOV1}--\eqref{TOV2}, starting from the stellar center and integrating outward until the radial pressure vanishes at the surface. The integration employs a standard fourth-order Runge-Kutta method with adaptively chosen step sizes to ensure numerical accuracy and convergence.

Our primary focus is understanding how two key parameters---the GB coupling $\alpha$ and the anisotropy parameter $\beta$---affect the mass-radius relation and overall structure of compact stars. The system of coupled first-order differential equations consists of Eq.~\eqref{TOV1} for the mass profile and Eq.~\eqref{TOV2} for the pressure gradient, supplemented by the modified Chaplygin gas equation of state and the quasi-local anisotropy prescription described in Sec.~\ref{sec:EOS}.

For the equation of state, we adopt $A = \sqrt{0.4}$ and $B = 0.23 \times 10^{-3}$ km$^{-2}$, keeping these fixed throughout our analysis. We then explore a parameter grid spanning $\alpha \in \{-10, -5, 0, 5, 0.5, 1.0\}$ km$^2$ and $\beta \in \{-1.0, -0.5, 0.0, 0.5, 1.0\}$. The choice $\alpha = 0$ recovers General Relativity, while $\beta = 0$ corresponds to isotropic matter. Negative values of $\alpha$ strengthen gravity compared to GR, whereas positive values weaken it. Similarly, $\beta$ controls the degree and sign of pressure anisotropy within the star.

For each parameter combination $(\alpha, \beta)$, we scan over a range of central densities $\rho_c$ spanning from $100$ MeV/fm$^3$ to $2000$ MeV/fm$^3$ to generate complete mass-radius sequences. The gravitational mass $M(R) = m(R)$ and radius $R$ of each configuration are recorded, along with the central density and compactness $M/R$. The results are presented in Figs.~\ref{fig3}--\ref{fig4} and Tables~\ref{table1}--\ref{table2}, demonstrating how the GB coupling and anisotropic pressure modify the stellar structure compared to standard general relativity.

\subsection{Mass-Radius and Compactness Relations: Dependence on the Gauss-Bonnet Coupling}

To understand the role of the GB coupling in decreasing monotonically the eristics and compactness in anisotropic dark energy stars, we explore the internal stellar architecture through radial profiles shown in Fig.~\ref{fig1}. This figure presents the energy density $\rho(r)$ alongside radial pressure $p_r(r)$ as functions of stellar radius for various $\alpha$ configurations. Both physical quantities follow a monotonically decreasing pattern from the center outward to the surface. Central values remain higher compared to boundary regions as $\alpha$ increases. The radial pressure adjusts accordingly to balance gravitational forces in the modified setting. Across all considered cases, the gravitational dynamics vanish smoothly at the stellar surface boundary, demonstrating numerical consistency. Table~\ref{table1} provides a summary of key numerical parameters employed in our calculations, and Fig.~\ref{fig3} shows mass--radius diagrams for equilibrium sequences across different $\alpha$ regimes.

In the limit where $\alpha = 0$, the formalism reduces precisely to General Relativity. Progressive increases in $\alpha$ from zero lead to weakening of the effective gravitational coupling via the modified field equations~\eqref{TOV1}--\eqref{TOV2}, permitting stellar configurations to sustain greater masses for given central density values.

 Positive $\alpha$ introduces higher-curvature corrections that counteract gravity, so mass-radius curves move towards more massive configurations. At the maximum value we explored ($\alpha = 10\,\text{km}^{2}$), our calculations show that the maximum mass exceeds GR predictions. This enhancement demonstrates that 4DEGB gravity allows for more massive compact objects and still retains its stability requirements. Negative $\alpha$ makes the gravitational interaction stronger and produces more compact stars with smaller radii and lower maximum masses.  In figure~\ref{fig3} we present our theoretical results and indicate the observational constraints imposed by the measurements of the massive pulsar PSR J0952--0607 ($M = 2.35 \pm 0.17\,M_{\odot}$)~\cite{Romani:2022jhd}, PSR J0740+6620 ($M = 2.08^{+0.07}_{-0.07}\,M_{\odot}$)~\cite{Fonseca:2021wxt} and PSR J0348+0432 ($M = 2.01 \pm 0.04\,M_{\odot}$)~\cite{Antoniadis:2013pzd}. The gravitational wave observations from GW170817~\cite{LIGOScientific:2018cki} and GW190814 (secondary object $2.50$--$2.67\,M_{\odot}$)~\cite{LIGOScientific:2020zkf} impose further constraints on the viable parameter space. Our sequences for $\alpha \geq 5\,\text{km}^{2}$ pass through these observational zones, providing evidence that 4DEGB models are in accord with astrophysical data.

In the context of relativistic stellar models, the compactness parameter, defined as $C = M/R$, provides a fundamental measure of gravitational strength within a self-gravitating configuration. General Relativity imposes a stringent upper limit on this quantity through the Buchdahl bound, which constrains stable isotropic spheres to $C < 4/9 \approx 0.444$. All equilibrium solutions obtained in our analysis comfortably satisfy this condition, confirming their dynamical viability. As the GB coupling parameter $\alpha$ increases, the maximum attainable compactness exhibits a mild reduction—an outcome consistent with the trend toward less dense, more extended stellar configurations. This behavior reflects the weakening of effective gravity in the positive--$\alpha$ regime. Overall, the adopted equation of state, when coupled with the 4DEGB corrections, produces stable, physically admissible stellar structures that satisfy compactness constraints and rely solely on the modified Chaplygin gas, without invoking any exotic matter sources.

\begin{figure}
    \centering
    \includegraphics[width = 8.5 cm]{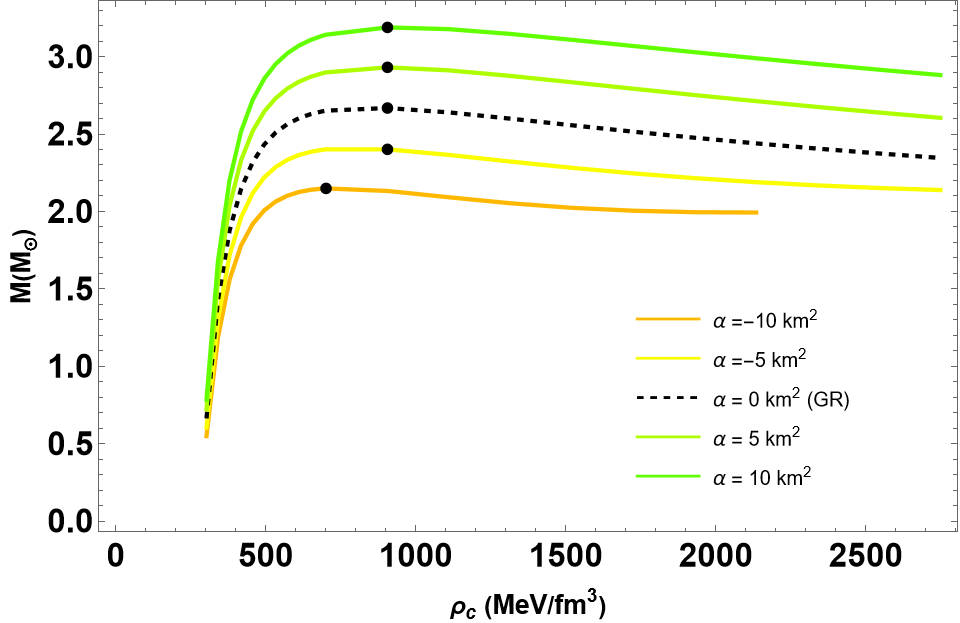}
    \includegraphics[width = 8.5 cm]{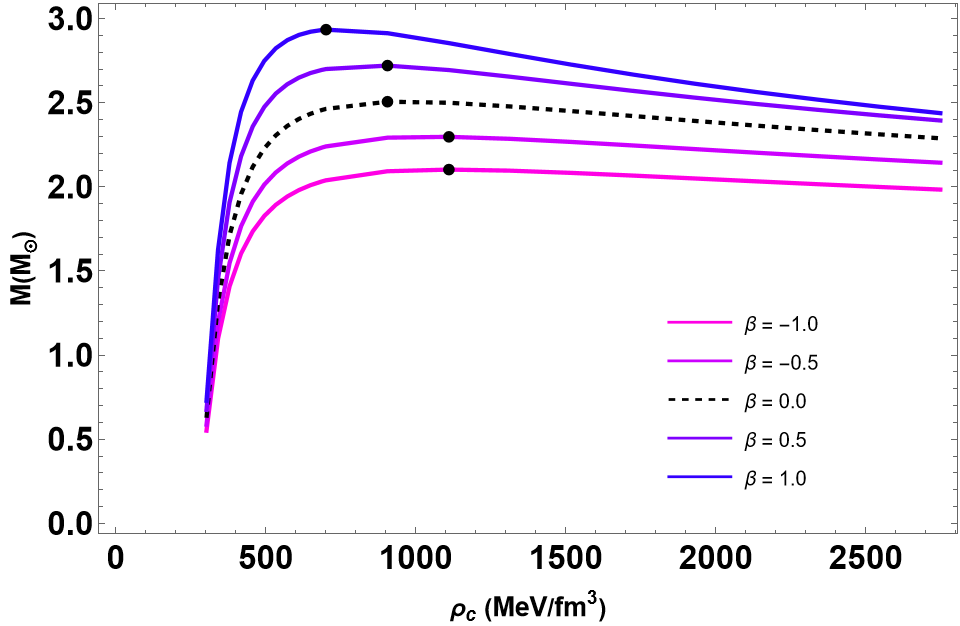}
    \caption{ Gravitational mass $M/M_{\odot}$ as a function of central energy density $\rho_c$ for anisotropic dark energy stars in 4DEGB gravity. \textbf{Left:} Sequences for varying GB coupling parameter $\alpha$. \textbf{Right:} Sequences for different anisotropy parameters $\beta$. Black dots denote maximum-mass configurations along each curve, marking the transition from stable to unstable stellar configurations according to the static stability criterion ($dM/d\rho_c < 0$). For consistency, the parameter sets listed in Tables~\ref{table1} and~\ref{table2} are employed in the present analysis.}
    \label{fig5}
\end{figure}

\begin{figure}
    \centering
    \includegraphics[width = 8.5 cm]{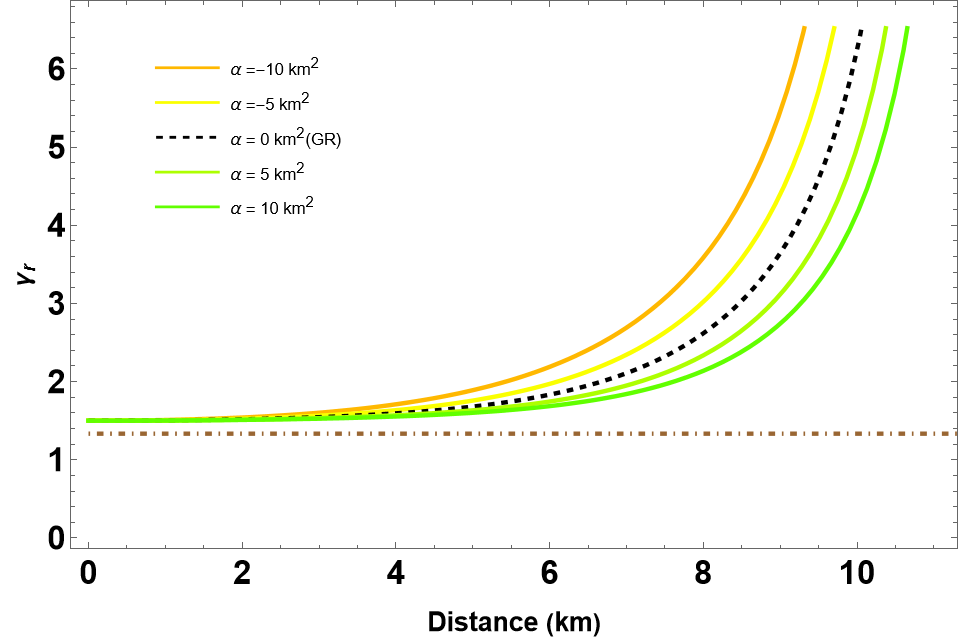}
    \includegraphics[width = 8.5 cm]{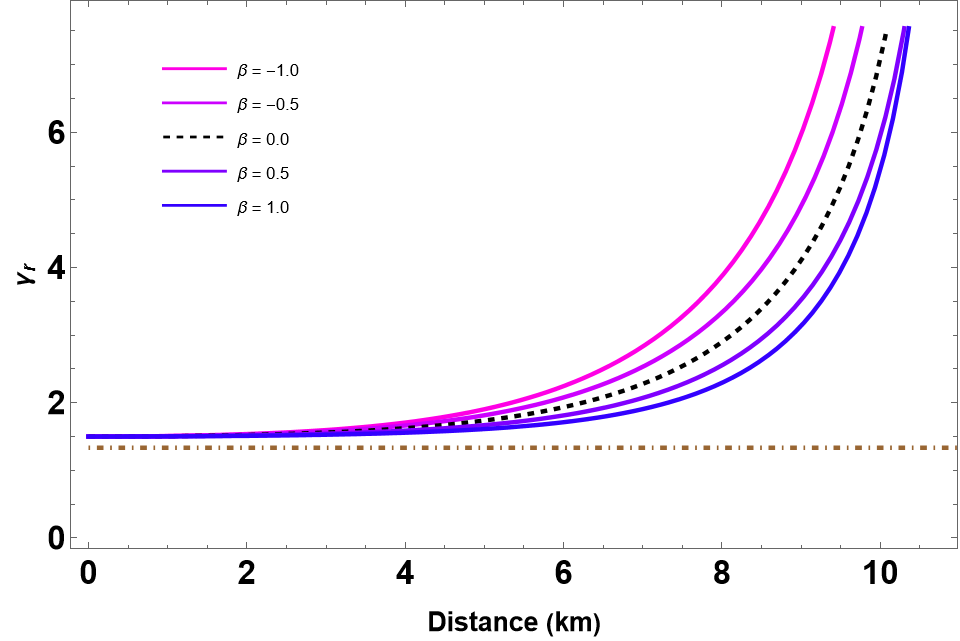}
    \caption{The radial adiabatic index ($\gamma_r$) is plotted as a function of the radial coordinate $r$ for anisotropic dark-energy stars in 4D~EGB gravity. \textbf{Left:} Profiles corresponding to different values of the GB coupling parameter $\alpha$. \textbf{Right:} Variation of $\gamma_r$ for distinct anisotropy parameters $\beta$. The same parameter sets as those used in Tables~\ref{table1} and~\ref{table2} are adopted for consistency.}
    \label{fig6}
\end{figure}

\subsection{Anisotropy Parameter Dependence: Impact of $\beta$ on Stellar Properties}

Having explored GB coupling effects, we now examine how pressure anisotropy shapes stellar structure. We fix $\alpha = 1.0\,\text{km}^{2}$ and vary $\beta$ over $[-1.0, +1.0]$, where negative values correspond to radially-dominated configurations ($p_r > p_t$) and positive values indicate tangential dominance ($p_t > p_r$). The case $\beta = 0$ represents the isotropic limit.

Figure~\ref{fig2} displays radial profiles of $\rho(r)$ and $p_r(r)$ for different $\beta$ values. Both quantities decrease monotonically from center to surface, maintaining regularity throughout. Unlike $\alpha$, which modifies the gravitational field equations directly, $\beta$ primarily influences pressure balance. Negative $\beta$ generates additional outward pressure support, while positive $\beta$ alters the stress distribution through tangential dominance.

Mass--radius relations in Fig.~\ref{fig4} exhibit systematic trends: increasing $\beta$ from $-1.0$ to $+1.0$ raises $M_{\max}$ from approximately $2.10\,M_\odot$ to $2.93\,M_\odot$, while radii expand from $11.32$ km to $12.50$ km (Table~\ref{table2}). This occurs because excess tangential pressure strengthens resistance to gravitational collapse. The left panel of Fig.~\ref{fig4} includes observational constraints from massive pulsars: 
PSR J0952--0607~\cite{Romani:2022jhd}, 
PSR J0740+6620 ~\cite{Fonseca:2021wxt}, and PSR J0348+0432 ~\cite{Antoniadis:2013pzd}. Also shown are the low-mass compact object HESS J1731--347~\cite{Doroshenko:2022nwp} and gravitational wave constraints from GW170817~\cite{LIGOScientific:2018cki} and the secondary component of GW190814~\cite{LIGOScientific:2020zkf}. Our predictions for $\beta \geq 0$ pass through these observational regions, with $\beta = +0.5$ and $+1.0$ aligning particularly well.

Compactness $\mathcal{C} = M/R$ increases moderately with $\beta$, rising from $\approx 0.276$ at $\beta = -1.0$ to $\approx 0.348$ at $\beta = +1.0$ (Fig.~\ref{fig4}, right panel). All configurations remain below the Buchdahl threshold $\mathcal{C} < 0.444$, confirming stability. Comparing Figs.~\ref{fig3} and~\ref{fig4} reveals that $\alpha$ and $\beta$ enhance maximum mass through distinct mechanisms: GB coupling via modified gravity, and anisotropy through pressure balance. Their combined effect can yield compact objects exceeding $3\,M_\odot$, consistent with the heaviest observed neutron star candidates.

\begin{figure}
    \centering
    \includegraphics[width = 8.5 cm]{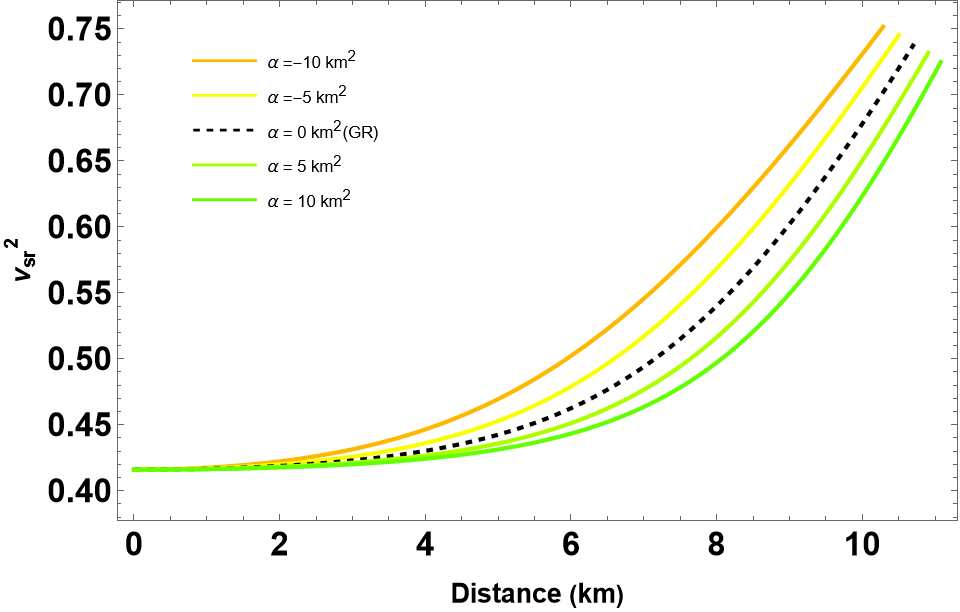}
    \includegraphics[width = 8.5 cm]{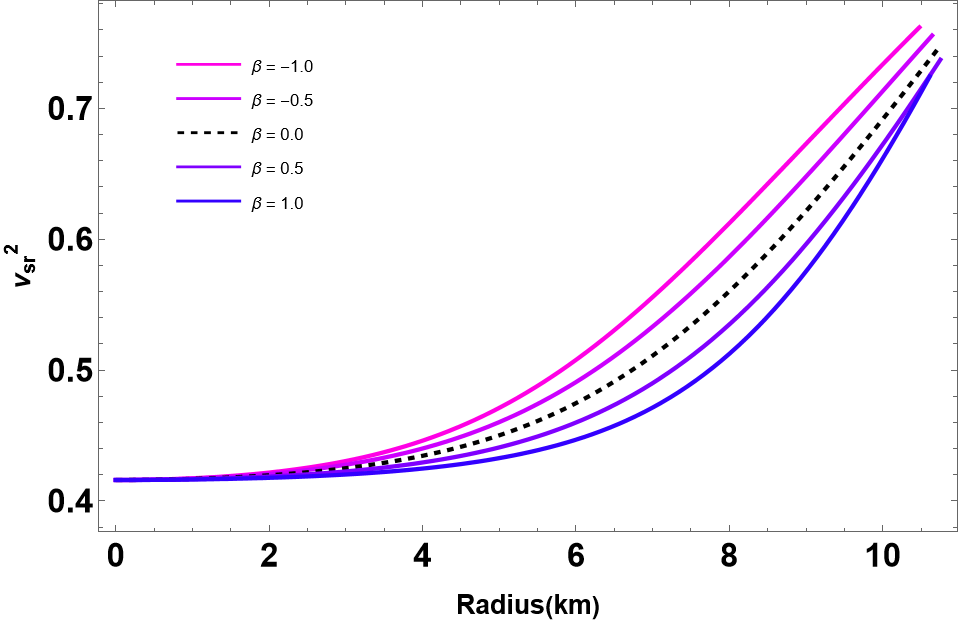}
    \includegraphics[width = 8.5 cm]{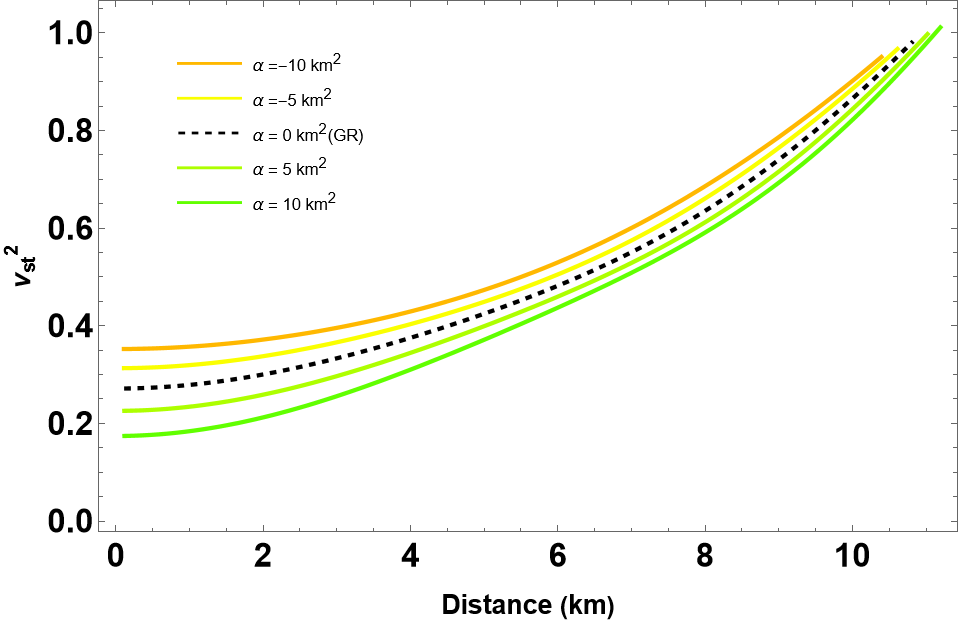}
    \includegraphics[width = 8.5 cm]{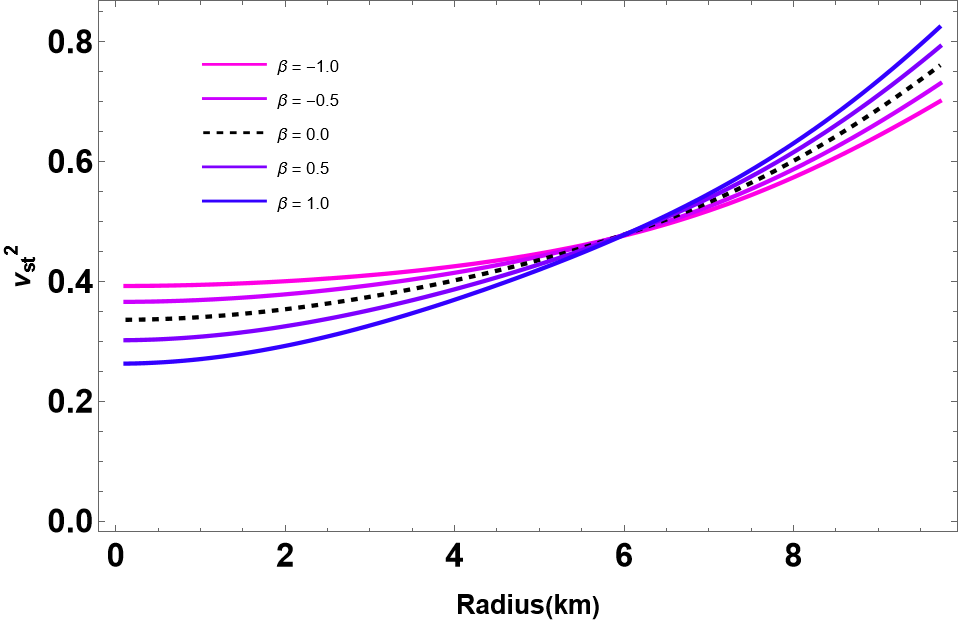}
    \caption{ Squared sound speeds in the radial ($v_r^2$, top panels) and transverse ($v_t^2$, bottom panels) directions as functions of the radial coordinate $r$ for anisotropic dark-energy stars in 4D~EGB gravity. \textbf{Left:} Profiles for different values of the GB coupling parameter $\alpha$. \textbf{Right:} Variation with the anisotropy parameter $\beta$. All models satisfy the causality condition $0 < v_r^2,\, v_t^2 \le 1$ throughout the stellar interior.}
    \label{fig7}
\end{figure}

\section{THE STATIC STABILITY CRITERION, ADIABATIC INDEX, AND SOUND VELOCITY}
\label{sec:stability}

 A comprehensive assessment of the equilibrium and stability of compact stellar configurations necessitates the concurrent evaluation of multiple diagnostic parameters. In the present analysis, particular attention is directed toward three fundamental stability indicators: (i) the static stability criterion inferred from the mass-central density relation, (ii) the adiabatic index governing the dynamical response to perturbations, and (iii) the propagation speeds of sound in both the radial and transverse directions. Collectively, these measures provide a consistent, physically transparent framework for assessing the structural integrity and stability of anisotropic dark-energy stars within the 4DEGB gravitational paradigm.

\subsection*{A. Static stability criterion}

A necessary condition for the stability of compact configurations is given by the classical turning-point, or static, stability criterion \cite{1965gtgc.book.....H,1971reas.book.....Z}. Along an equilibrium sequence generated by varying the central energy density $\rho_c$, stable configurations correspond to the branch where
\begin{equation}
\frac{dM}{d\rho_c} > 0,
\end{equation}
while the onset of instability occurs once the maximum-mass configuration is reached and the slope changes sign,
\begin{equation}
\frac{dM}{d\rho_c} < 0.
\end{equation}
The points of maximum mass, indicated by black dots in Fig.~\ref{fig5}, therefore mark the boundary between stable and unstable regions. Although this condition is necessary but not sufficient for full dynamical stability, it offers a reliable criterion for assessing the behavior of equilibrium sequences in the absence of explicit perturbation analysis. We find that increasing the GB coupling parameter $\alpha$ or introducing positive anisotropy ($\beta>0$) generally shifts the maximum-mass point toward higher $\rho_c$, implying an extended stable region in parameter space.

\subsection*{B. Radial adiabatic index}

An important indicator of the dynamical stability of a self-gravitating configuration is the adiabatic index, denoted by $\gamma_r$. This quantity measures the stiffness of the stellar fluid and determines how the pressure responds to small, adiabatic perturbations in density. The concept was first introduced by Chandrasekhar~\cite{Chandrasekhar:1964zz} through a variational formulation of the pulsation equations for relativistic stars, which established that a stable configuration must satisfy a lower limit on $\gamma_r$ throughout its interior.

Following this seminal approach, the adiabatic index has been widely employed to examine the stability of compact objects such as neutron stars and quark stars within this gravity context~\cite{Pretel:2025roz,Banerjee:2020dad,Gammon:2023uss}. In the present framework, the radial adiabatic index is defined as
\begin{equation}\label{adi_new}
    \gamma_r(r) = \left(1 + \frac{\rho}{p_r}\right) \left(\frac{d p_r}{d\rho}\right)_{S},
\end{equation}
where $p_r$ is the radial pressure, $\rho$ is the energy density, and the derivative is taken at constant entropy ($S$). The factor $d p_r/d\rho$ represents the square of the radial sound speed, ensuring that $\gamma_r$ encapsulates both the compressibility and the inertial response of the stellar matter.

In Newtonian gravity, the critical adiabatic index for marginal stability is $\gamma_{\mathrm{cr}} = 4/3$. When relativistic effects are included, this threshold increases slightly because stronger gravitational fields enhance the tendency toward collapse~\cite{Moustakidis:2016ndw}. Moreover, anisotropy in the fluid introduces additional pressure gradients that modify the local equilibrium conditions, often resulting in $\gamma_{\mathrm{cr}} > 4/3$~\cite{Nashed:2022zxm,Sharif:2020xwm}. Therefore, a physically acceptable stellar model must satisfy $\gamma_r(r) > \gamma_{\mathrm{cr}}$ everywhere inside the configuration.

Our numerical analysis, illustrated in Fig.~\ref{fig6}, shows that $\gamma_r$ remains greater than $4/3$ throughout the interior of the dark-energy star for all viable choices of the GB coupling $\alpha$ and anisotropy parameter $\beta$. Near the stellar center, $\gamma_r$ attains its minimum value but still exceeds the critical limit, while toward the surface it increases monotonically as the density and pressure decrease. This monotonic behavior indicates that the configurations are dynamically stable under infinitesimal radial oscillations. Furthermore, an increase in either $\alpha$ or $\beta$ enhances the magnitude of $\gamma_r$, implying that both higher-curvature corrections and tangential stresses contribute positively to the stiffness of the matter distribution and thus to the overall stability of the star.

\subsection*{C. Sound velocity and causality condition}

Another essential aspect of the stability analysis is the examination of the propagation of infinitesimal perturbations through the stellar fluid. The squared sound speeds in the radial and transverse directions are expressed as
\begin{equation}
v_r^2 = \frac{d p_r}{d \rho}, \qquad 
v_t^2 = \frac{d p_t}{d \rho} = \frac{d(p_r + \sigma)}{d\rho} = v_r^2 + \frac{d\sigma}{d\rho},
\end{equation}
where $\sigma = p_t - p_r$ denotes the pressure anisotropy. The physical requirement of causality demands that both the radial and transverse sound speeds remain subluminal, i.e.,
\begin{equation}
0 \le v_r^2,\, v_t^2 \le 1,
\end{equation}
so that pressure disturbances propagate within the light cone throughout the stellar interior.

The profiles of $v_r^2$ and $v_t^2$, presented in Fig.~\ref{fig7}, satisfy this causality condition for all equilibrium configurations considered in the present study. Both quantities increase gradually from the stellar center toward the surface, reflecting the stiffening of the equation of state as the density decreases. No violation of the causality limit is observed for any value of the GB coupling $\alpha$ or the anisotropy parameter $\beta$. 

The overall behavior of the sound speeds is consistent with the conclusions drawn from the adiabatic index and the turning-point stability criterion: configurations that fulfill the causal bound also exhibit $\gamma_r > 4/3$ and lie on the stable branch of the $M(\rho_c)$ relation. This agreement among independent stability diagnostics provides strong evidence that the anisotropic dark-energy stars analyzed within 4DEGB gravity remain dynamically stable under small radial perturbations within the considered parameter domain. The numerical results presented in Figs.~\ref{fig5}--\ref{fig7} and Tables~\ref{table1}--\ref{table2} collectively support this conclusion.

\section{Conclusions}
\label{sec:conclusions}

In this work we have constructed and analyzed equilibrium models of anisotropic dark-energy compact stars in the regularized 4D Einstein–Gauss–Bonnet (4DEGB) gravity. The interior matter is modeled by the modified Chaplygin gas equation of state, and pressure anisotropy is introduced through a quasi-local ansatz proportional to the local compactness. By numerically integrating the modified TOV system across a broad grid of central densities, Gauss–Bonnet couplings $\alpha$, and anisotropy parameters $\beta$, we have explored the resulting mass-radius relations, internal structure, and stability properties.

Our principal findings are as follows. Positive values of the GB coupling ($\alpha>0$) act effectively to weaken the gravitational pull compared with the GR limit, thereby permitting more massive and less compact equilibria for the same central density. Tangentially dominated anisotropy ($\beta>0$) similarly raises the maximum mass and expands stellar radii by providing extra support against gravitational collapse. When combined, these effects produce sequences that can exceed the conventional $2\,M_\odot$ threshold and reach the upper end of the compactness parameter space while remaining within the appropriate 4DEGB stability bounds.

We have tested the dynamical robustness of the resulting configurations using three complementary diagnostics. The turning-point criterion applied to $M(\rho_c)$ sequences identifies the secular stability boundary and the maximum-mass points; the radial adiabatic index $\gamma_r(r)$ remains greater than the Newtonian benchmark $4/3$ throughout the interior for the viable models considered, indicating strong local resistance to radial collapse. The squared sound speeds $v_r^2$ and $v_t^2$ are everywhere subluminal in our numerical solutions, ensuring microphysical causality. Fundamental-mode eigenfrequencies computed for representative models corroborate the turning-point classification, confirming that models on the stable branch resist linear radial perturbations. Taken together, these independent checks provide consistent evidence that a broad portion of the $(\alpha,\beta)$ plane admits dynamically stable anisotropic dark-energy stars.

We have also confronted our theoretical predictions with recent observational constraints from gravitational-wave events (GW170817, GW190814) and precision pulsar mass measurements (e.g., PSR J0952–0607, PSR J0740+6620). This comparison delimits the phenomenologically viable parameter region: moderate positive $\alpha$ and nonzero tangential anisotropy yield models that naturally accommodate the heaviest observed compact objects while respecting causality and compactness limits. Notably, some equilibrium sequences produce ultra-compact configurations that approach, but do not violate, the modified Buchdahl bound in 4DEGB gravity; such extreme compact objects (ECOs) are observationally distinguishable from classical black holes in principle and therefore merit further study.

Finally, we summarize several caveats and future avenues. Our models rely on a phenomenological MCG equation of state and a quasi-local anisotropy prescription; improved microphysical modeling (e.g., matching to microscopic interaction models or including rotation and magnetic fields) would refine quantitative predictions. Extending the perturbative stability analysis to include nonradial modes, full general-relativistic time evolutions, and a systematic parameter inference using Bayesian comparison to multi-messenger data are clear next steps. Observationally, targeted searches for signatures of ECOs or deviations in tidal deformabilities and moment of inertia may provide the most direct tests of the higher-curvature and anisotropy effects discussed here. Overall, our results indicate that anisotropic dark-energy stars in 4DEGB gravity are a viable and testable class of compact-object solutions, motivating continued theoretical and observational scrutiny.

\begin{acknowledgments}
T. Tangphati acknowledges COST action CA21106 and CA22113.
A. Pradhan gratefully acknowledges the support and resources provided by the Inter-University Centre for Astronomy and Astrophysics (IUCAA), Pune, India, under its Visiting Associateship Programme.

\end{acknowledgments}

\bibliography{References}

\end{document}